\documentclass[aps,pra,superscriptaddress,twocolumn]{revtex4-2}
\usepackage{amsmath}
\usepackage{graphicx}	
\usepackage{bm} 
\usepackage{color}
\usepackage{setspace}
\usepackage{placeins}
\usepackage{natbib}
\usepackage[normalem]{ulem}
\usepackage{hyperref}
\hypersetup{colorlinks=true,citecolor=blue,linkcolor=blue,urlcolor=blue}
\begin{document}
\title{Resonant superfluidity in the Rabi-coupled spin-dependent Fermi-Hubbard model}
\date{\today}
\author{Mathias~Mikkelsen}
\email[]{mathias-mikkelsen@phys.kindai.ac.jp}
\affiliation{Department of Physics, Kindai University, Higashi-Osaka City, Osaka 577-8502, Japan}
\author{Ryui~Kaneko}
\affiliation{Department of Physics, Kindai University, Higashi-Osaka City, Osaka 577-8502, Japan}
\author{Daichi~Kagamihara}
\affiliation{Department of Physics, Kindai University, Higashi-Osaka City, Osaka 577-8502, Japan}
\author{Ippei~Danshita}
\email[]{danshita@phys.kindai.ac.jp}
\affiliation{Department of Physics, Kindai University, Higashi-Osaka City, Osaka 577-8502, Japan}
\begin{abstract} 

We investigate the ground-state phase diagram of the one-dimensional attractive Fermi-Hubbard model with spin-dependent hoppings and an on-site Rabi coupling using the density matrix renormalization group method. In particular, we show that even in the limit of one component being immobile the pair superfluidity can be resonantly enhanced when the Rabi coupling is on the order of the interaction strength just before the system starts to strongly polarize. We derive an effective spin-1/2 XXZ model in order to understand the ground-state properties in the strong attraction limit.  
\end{abstract}

\maketitle
\section{Introduction}

Thanks to their high controllability, systems of ultracold two-component Fermi gases have served as an ideal playground for studying superfluids (SF) consisting of fermions \cite{Giorgini2008,Randeria2014}. Since the realization of the crossover between a Bardeen-Cooper-Schrieffer (BCS) type SF and Bose-Einstein condensate (BEC) \cite{Regal2004,Zwierlein2004}, which is caused by tuning the s-wave scattering length using a Feshbach resonance \cite{Chin2010}, various types of phenomena related to superfluidity have been observed, such as elementary excitations \cite{Hoinka2017,Behrle2018}, quantum vortices \cite{Zwierlein2005}, and 
SF critical velocities \cite{Miller2007,Weimer2015}. 

While most experimental studies on two-component Fermi gases with attractive interactions have been performed in continuum systems, some experiments have used optical lattices in order to address lattice systems \cite{Strohmaier2007,Paredes2010,Mitra2018,Brown2020}, which are described by the attractive Fermi-Hubbard model (FHM) \cite{Micnas1990}. Since the temperatures realized in the lattice systems so far are above the SF phase transition temperature, the precursors of SF phase transitions, namely the development of pairing correlations \cite{Mitra2018} and the pseudo gap in the occupied spectral function \cite{Brown2020}, have been investigated.  

The use of optical lattices further improves the controllability of the system. For instance, the single-particle dispersion relation can be changed rather flexibly since several types of lattice geometry, including cubic \cite{Greiner2002}, chain \cite{Tilman2004}, ladder \cite{Chen2011}, square \cite{Spielman2007}, triangular \cite{Becker2010}, honeycomb \cite{Tarruell2012}, kagome \cite{Jo2012}, and Lieb \cite{Taie2015} lattices, have been realized. Moreover, state-dependent optical lattices \cite{Gadway2011,Riegger2018,Ono2021} allow for controlling the ratio between the hoppings of the two components. While a rich phase diagram emerges as a function of the hopping imbalance and the interaction strength \cite{Cazalilla2005}, it is well known that superfluidity is suppressed in general for a large hopping imbalance. In the limit that one component is completely immobile, where the model is reduced to the Falicov-Kimball model \cite{Falicov1969}, the SF state is not present.

In this paper, we add an on-site Rabi coupling between the two components to the attractive Hubbard model with unequal hoppings in order to investigate its effects on superfluidity. Liu {\it et al.}~\cite{Liu2004} have introduced such a model and analyzed it within a mean-field approximation. The fact that the Rabi coupling  has been successfully implemented and controlled in recent experiments using optical lattices loaded with ultracold gases \cite{Krinner2018,Riegger2019} has led to renewed theoretical interest in understanding the physics of such variations of the attractive FHM. We solve the model at one spatial dimension numerically utilizing the density matrix renormalization group (DMRG) method \cite{White1992} in order to calculate the ground-state properties as a function of the Rabi coupling and the on-site attraction. In the case that the hopping of one component is zero suppressing superfluidity, we show that the superfluidity can be resonantly restored when the Rabi coupling is slightly smaller than the attractive interaction strength. We explain these results at large attractive interactions in terms of an effective spin-1/2 XXZ model. 

The remainder of the paper is organized as follows. In section \ref{sec:model} we introduce the FHM under investigation and explain how a unitary rotation clarifies the role of the model parameters. In section \ref{sec:results} we present our results, first introducing the relevant physical correlation functions and their significance in section \ref{sec:CorrelationfunctionQLRO}. In section \ref{sec:DMRGresulthalffilling} we investigate the commensurate half-filling case utilizing DMRG. In section \ref{sec:EffectiveHamiltonian} we introduce an effective spin-1/2 XXZ model for large attractive interactions and discuss how it explains the results of \ref{sec:DMRGresulthalffilling}. In section \ref{sec:SchematicPhaseDiagram} we present a schematic phase diagram based on the effective model as well as DMRG results for the correlations of the full model at some representative parameters. In section \ref{sec:SuperfluidCDWfilling} we further investigate the validity of the schematic phase diagram in the full FHM by means of DMRG calculations. Specifically, we investigate the correlations as a function of the number of particles per site. Section \ref{sec:summary} contains a summary and outlook. The appendix \ref{sec:appendix} contains a detailed derivation of the effective spin-1/2 XXZ model based on second-order degenerate perturbation theory.

\section{Model}
\label{sec:model}

We consider an ultracold two-component Fermi gas in a three-dimensional optical lattice. We assume that the optical lattice is so deep in the transverse directions (say, $xy$ directions) that the motion of the particles is forbidden in these directions, i.e., the system is one-dimensional. We anticipate that the two components correspond to either two different Zeeman sublevels of an alkali atom or two internal states of electrons in an alkali-earth(-like) atom. We regard the two components as the (pseudo-)spin degrees of freedom in the FHM. We also assume that there exists an on-site Rabi coupling between the two components, which can be correspondingly created by either near-resonant radio frequency radiation or a laser. When the optical lattice in the longitudinal direction ($z$ direction) is component-dependent and sufficiently deep, the system is well described by the attractive FHM with an on-site Rabi coupling and component-dependent hoppings~\cite{Liu2004},
\begin{align}
\hat{H} &=
\sum_{\sigma,j}
\Big(
-t_{\sigma} \hat{c}_{\sigma,j}^\dagger \hat{c}_{\sigma,j+1} + H.C. \Big)
- U\sum_{j} \hat{n}_{1,j} \hat{n}_{2,j} \nonumber\\
&+\sum_{j}
\Big[
\Big(-\mu-\frac{\delta}{2}\Big) \hat{n}_{1,j} + \Big(-\mu+\frac{\delta}{2}\Big) \hat{n}_{2,j}
\Big]
\nonumber \\
&+\frac{\Omega_0}{2} \sum_{j} (\hat{c}_{1,j}^\dagger \hat{c}_{2,j}+\hat{c}_{2,j}^\dagger \hat{c}_{1,j}).
\label{eq:BasicHubbardHamiltonian}
\end{align}
Here $\hat{c}_{\sigma,j}^{\dagger}$, $\hat{c}_{\sigma,j}$, and $\hat{n}_{\sigma,j} = \hat{c}_{\sigma,j}^{\dagger}\hat{c}_{\sigma,j}$ are the creation, annihilation, and number operators of spin $\sigma(=1,2)$ at site $j$, $t_{\sigma}$ is the hopping of spin $\sigma$ between nearest-neighboring sites, $U$ is the strength of the on-site inter-particle attraction, $\mu$ is the chemical potential, and $\Omega_0$ and $\delta$ are the strength of the Rabi coupling and detuning. All the parameters can be widely controlled in actual experiments. For convenience, we hereafter call the model of Eq.~(\ref{eq:BasicHubbardHamiltonian}) the Rabi-coupled spin-dependent FHM (RSFHM).

\subsection{On-site diagonalization and rotated frame}
The RSFHM represented in terms of $\hat{c}_{\sigma,j}^{\dagger}$ and $\hat{c}_{\sigma,j}$ as in 
Eq.~(\ref{eq:BasicHubbardHamiltonian}) is physically meaningful in the sense that the component index $\sigma$ indeed corresponds to two internal states of an atom. However, some of the essential properties of the model become clearer by considering a unitary rotation of the Hamiltonian with an angle which diagonalizes the on-site Hamiltonian at a given value of $\delta$ and $\Omega_0$. The unoccupied and doubly occupied states ($| 0 \rangle $ and $| 2 \rangle =\hat{c}^{\dag}_2 \hat{c}^{\dag}_1 | 0 \rangle$) are eigenstates of this Hamiltonian with eigenvalues 0 and $-U$ respectively, while the two remaining eigenstates are given by the creation operators $\hat{c}_{+}^\dagger= \alpha \hat{c}_{1}^\dagger+\beta \hat{c}_{2}^\dagger$ and $\hat{c}_{-}^\dagger= -\beta \hat{c}_{1}^\dagger+\alpha \hat{c}_{2}^\dagger$, where $\alpha=\cos(\theta),\beta=\sin(\theta)$. Here $\theta$ is an angle determined by the values of $\delta$ and $\Omega_0$ by the equations  $\alpha \beta= \frac{\Omega_0}{2\sqrt{\Omega_0^2+\delta^2}}$ and $\beta^2-\alpha^2= \frac{\delta}{\sqrt{\Omega_0^2+\delta^2}}$. The corresponding on-site eigenvalues are given by 
\begin{align}
E^{\pm} = -\mu \pm\frac{1}{2} \sqrt{\delta^2+\Omega_0^2}.
\end{align}

Rotating the full Hamiltonian at the angle $\theta$, it can be written in terms of the operators $\hat{c}_{+,j}$ and $\hat{c}_{-,j}$ as 
\begin{align}
\hat{H}_{\theta} &= \sum_{\zeta=+,-}\sum_{j} \Big(
-t_{\theta,\zeta}  \hat{c}_{\zeta,j}^\dagger \hat{c}_{\zeta,j+1} + H.C.
\Big)
\nonumber \\
&+\gamma_\theta \sum_{j}
\Big(
\hat{c}_{+,j}^\dagger \hat{c}_{-,j+1}+\hat{c}_{-,j}^\dagger \hat{c}_{+,j+1} + H.C.
\Big)
\nonumber \\
+&\sum_{j}
\Big[
(-\mu+\delta_{\theta})\hat{n}_{+,j}+ (-\mu-\delta_{\theta})\hat{n}_{-,j}
\Big]
-U\sum_{j}\hat{n}_{+,j}\hat{n}_{-,j}
\label{eq:RotatedHubbardHamiltonian}
\end{align}
with
\begin{align*}
&t_{\theta,+} = t_{1} \alpha^2+t_{2} \beta^2   \quad , \quad \gamma_{\theta} = \frac{\Omega_0}{2\sqrt{\Omega_0^2+\delta^2}}(t_1-t_2)  \\
& t_{\theta,-} = t_{2} \alpha^2+t_{1} \beta^2  \quad , \quad  \delta_{\theta} = \frac{1}{2}\sqrt{\Omega_0^2+\delta^2}. 
\end{align*}  
The interaction and chemical potential terms are covariant under this transformation, while the $\delta$ and $\Omega_0$ fields combine as a Zeeman-type field for the new Hamiltonian with the direction determined by the angle $\theta$. For $t_1=t_2$ the rotated Hamiltonian is just the FHM with a Zeeman field that has a symmetric contribution from the Rabi and Zeeman terms of the original Hamiltonian. The SU(2) symmetry, which is present before the application of the on-site fields, remains such that the physics is essentially the same regardless of the orientation of the field. This model has been investigated in the literature, famously leading to the emergence of a Fulde–Ferrell–Larkin–Ovchinnikov~(FFLO) phase between the BCS phase at small fields and the fully polarized phase at large fields \cite{Fulde1964,Larkin1964}. See also \cite{Rizzi2008,Luscher2008} for exact numerical studies of the FFLO phase in a population-imbalanced one-dimensional FHM. 

On the other hand, since $t_1\neq t_2$ breaks the initial SU(2) symmetry, the orientation of the on-site field becomes important. For $\Omega_0=0$ the rotated Hamiltonian corresponds to the original Hamiltonian, that is, a spin-dependent Hubbard model with a Zeeman field $\delta$ which has been studied in, e.g., \cite{Wang2009}. This case has been investigated in order to understand how the FFLO phase is affected by asymmetric hopping. For $\Omega_0\neq 0$ the effective hoppings of the rotated components change but are still asymmetric and an extra symmetric hopping term which transforms between the components is added. This means that $[\hat{H}_\theta,\hat{n}_{\pm,j}] \neq 0$ breaking the U(1) symmetry of these components, unlike the $t_1=t_2$ or $\Omega_0=0$ case. So even if the system is defined in the rotated frame, the Hilbert space cannot be divided into separate spin sectors. On the other hand, the system conserves the total number of particles which is therefore still a good quantum number. 

\section{Results}
\label{sec:results}

The main focus of this investigation is the limiting case $t_2=0$ where superfluidity is entirely suppressed in the absence of the Rabi coupling. It therefore provides the clearest setting in which to study the effect of the Rabi coupling on superfluidity.  As we keep the total particle number fixed, the chemical potential is unimportant for our results and we set it to zero. We utilize the C++ version of the ITensor library \cite{Fishman2020} in order to implement a standard DMRG algorithm. In general, we keep the truncation dimension capped at 200 which leads to well-converged results. However, due to the degeneracy of the ground states in the $\Omega_0=0$ limit, it is difficult to obtain correct results for small values of $\Omega_0$ where low-lying states are almost degenerate and the DMRG algorithm tends to end up in a superposition of almost degenerate ground states. At half filling, well-converged results corresponding to the true ground state are generally obtained for $\Omega_0>0.1 U$, while larger values of $\Omega_0$ are required away from half filling. 
\par
We therefore investigate the half-filling case in section \ref{sec:DMRGresulthalffilling}. Motivated by findings in the half-filling case, we introduce an effective theory in the strongly attractive limit in section \ref{sec:EffectiveHamiltonian}. From this, we give a more detailed explanation of the physics observed in the paired regime. The effective model is valid for any combination of parameters $t_1,t_2,\Omega_0,\delta$ (as long as $U$ is large), so that it becomes clear that the physics away from the limiting case of $t_2=0$ is not substantially different. In section \ref{sec:SchematicPhaseDiagram}, we therefore suggest a schematic phase diagram for the system at $t_2=0$ as a function of $\Omega_0$ and the number of particles per site $n=N/L$ based on the effective model and DMRG calculations for the full model at some representative parameters.  Finally, we further probe the validity of the schematic phase diagram in the full model for strong and moderate attraction by varying the number of particles per site $n$ in a DMRG calculation in section \ref{sec:SuperfluidCDWfilling}. In order to investigate these features we first define how to quantify the SF and charge-density-wave~(CDW) order.

\subsection{Correlation functions and off-diagonal quasi-long-range order}
\label{sec:CorrelationfunctionQLRO}
For attractive interactions, quasi-long-range order (QLRO) of the pairing and/or CDW can be formed in the ground states of the RSFHM  model. The properties of the pairs and density waves can be probed by the correlation functions 
\begin{align}
P(j,k) &=  \langle \hat{c}_{+,j}^\dagger \hat{c}_{-,j}^\dagger \hat{c}_{-,k} \hat{c}_{+,k}  \rangle = \langle \hat{c}_{1,j}^\dagger \hat{c}_{2,j}^\dagger \hat{c}_{2,k} \hat{c}_{1,k}  \rangle \\
C(j,k) &=  \langle (\hat{n}_{+,j}+\hat{n}_{-,j}-\bar{n}) (\hat{n}_{+,k}+\hat{n}_{-,k}-\bar{n})  \rangle \nonumber \\
&=\langle (\hat{n}_{1,j}+\hat{n}_{2,j}-\bar{n}) (\hat{n}_{1,k}+\hat{n}_{2,k}-\bar{n})  \rangle. \label{eq:densitycorrelations}
\end{align}
Here $\bar{n}$ is the average density and is subtracted in order to remove the trivial lattice ordering. Note that both functions are invariant under the rotation of the Hamiltonian and can therefore be calculated in either frame, using the relevant creation and annihilation operators.  The numerical calculations in this manuscript are performed in the original frame of the Hamiltonian
given in Eq.~\eqref{eq:BasicHubbardHamiltonian}. In accordance with general convention for one-dimensional systems (see, e.g., \cite{Rigol2005,Ueda2008}), we refer to the presence of the off-diagonal QLRO, signified by a power-law decay of $P(|j-k|)$, as a SF (quasi-)order throughout the manuscript.  The density-density correlations $C(j,k)$ can similarly display off-diagonal QLRO corresponding to a CDW (quasi-)order. The model under consideration can also display true long-range CDW order depending on the parameters.

A simple way to gauge the degree of pairing is the average number of pairs $n_{\text{pair}}=2 N_{\text{pair}}/N$ where $N$ is the number of particles and
\begin{align}
N_{\text{pair}}=\sum_{j} P(j,j)
\end{align}
while polarization along the $\theta$ direction
\begin{align}
M &= \sum_{j} \langle \hat{S}^z_j \rangle = \sum_{j} \langle \hat{n}_{+,j}-\hat{n}_{-,j}\rangle \nonumber \\
&= \sum_{j} \frac{\delta}{\sqrt{\delta^2+\Omega_0^2}}\langle \hat{n}_{2,j}-\hat{n}_{1,j}\rangle \nonumber\\
&+\frac{\Omega_0}{\sqrt{\delta^2+\Omega_0^2}}[\hat{c}_{1,j}^\dagger \hat{c}_{2,j}+\hat{c}_{2,j}^\dagger \hat{c}_{1,j}],
\end{align}
quantifies the imbalance between $+$ and $-$ particles.

The presence of QLRO in the real-space correlation functions can also be probed by the eigenvalues of $P(j,k),C(j,k)$ or the Fourier transforms of $P(j,k),C(j,k)$, see, e.g., \cite{Yang1962,Rigol2005,Ueda2008}. The Fourier transforms of $P(j,k)$ and $C(j,k)$ correspond to the pair momentum distribution and static structure factor respectively
\begin{align}
n(k) &= \frac{1}{L} \sum_{m,n}e^{-i(m-n)k} P(m,n) \\
S(\nu) &=\frac{1}{L} \sum_{m,n}e^{-i(m-n)\nu} C(m,n), 
\end{align}
where $L$ is the number of sites, and these are of particular interest as both are experimentally measurable in cold atom setups \cite{Mitra2018} allowing for the experimental determination of QLRO. If the system is homogeneous and the boundary condition is periodic, the peaks in Fourier space are equivalent to the maximum eigenvalues of the real-space correlation functions, which can be diagonalized as $P(j,k)/N_{\text{pair}}=\sum_{n}\lambda_{P,n} \phi_{P,n}(j)\phi_{P,n}(k)$ and $C(j,k)/\sum_{j} C(j,j)=\sum_{n}\lambda_{C,n} \phi_{C,n}(j)\phi_{C,n}(k)$, where $\phi_{P(C),n}(j)$ are the eigenfunctions of a pair (density) correlation function. We denote these eigenvalues as $\lambda_{P,0}$ and $\lambda_{C,0}$ for the SF and CDW correlations respectively. Even in the case of an open boundary condition, which we adopt in our DMRG analyses shown below, the maximal values of $n(k)/N_{\text{pair}}$ and $S(\nu)/\sum_{j} C(j,j)$ are well approximated by $\lambda_{P,0}$ and $\lambda_{C,0}$. 

The fact that maximum eigenvalues are proportional to $L^a$ (with $0<a<1$) while the remaining eigenvalues do not scale with system size indicates the presence of QLRO in the system \cite{Yang1962}. If the maximum eigenvalues are constant with system size, however, it indicates the absence of order. In order to test this rigorously and extrapolate the eigenvalues to the thermodynamic limit, one must investigate  $\lambda_{P,0}N_{\text{pair}}$ or $\lambda_{C,0}\sum_{j} C(j,j)$ as a function of the system size. For constant $L$, however, a large value of $\lambda_{P,0}$ or $\lambda_{C,0}$ compared to the other eigenvalues signifies QLRO in the real-space correlation function at that system size. With the above normalization, the sum of eigenvalues is unity such that $\lambda_{P,0}$ and $\lambda_{C,0}$ correspond to the relative size of the largest eigenvalues and can measure the QLRO. Comparing $\lambda_{P,0}$ and $\lambda_{C,0}$  also allows us to probe the dominant order at a constant $L$ as the same normalization is employed for both.

\subsection{DMRG results at half filling for $t_2=0$}
\label{sec:DMRGresulthalffilling}

\begin{figure*}
\centering
\includegraphics[width=0.9\linewidth]{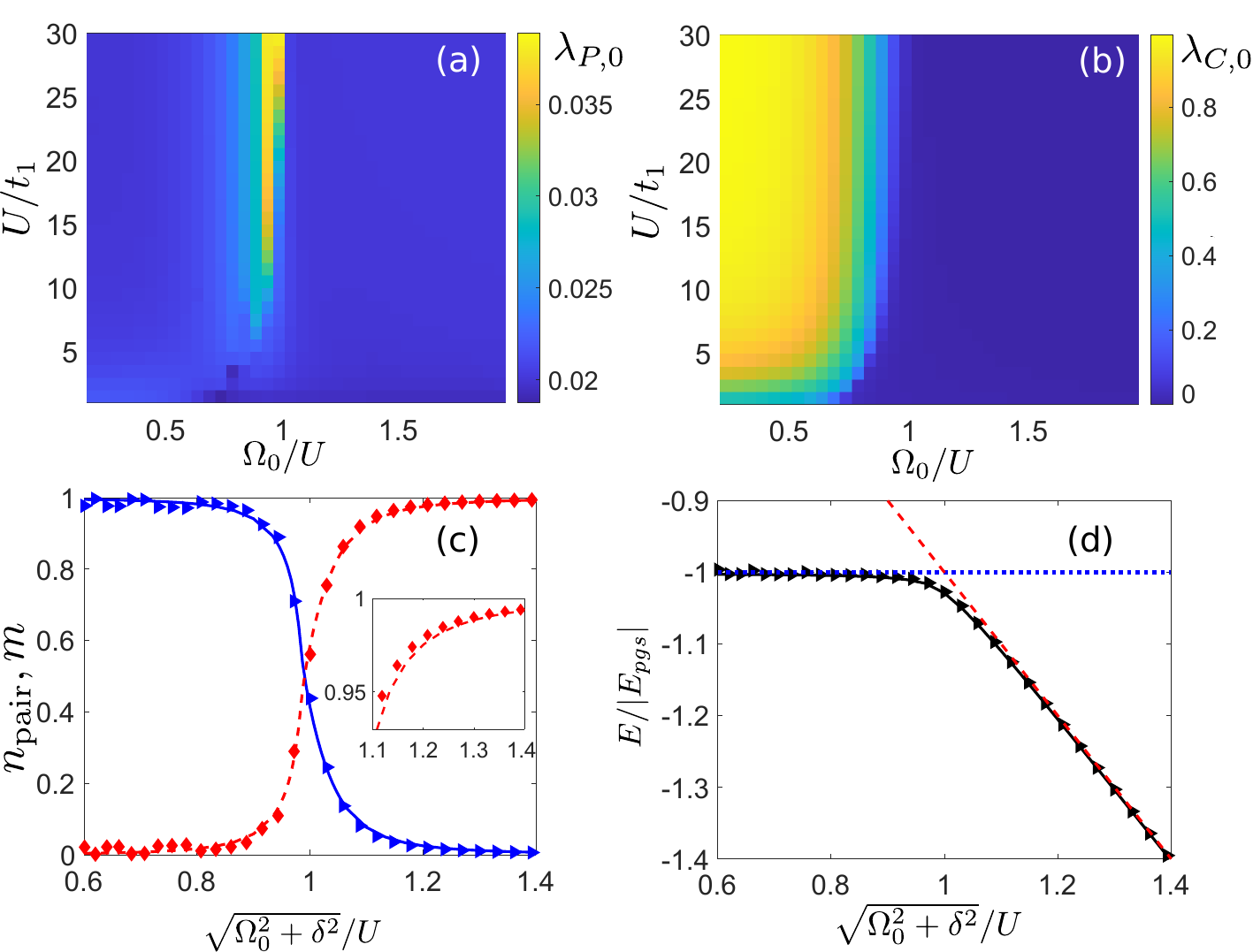}
\caption{(a) $\lambda_{P,0}$ and (b) $\lambda_{C,0}$, which measure the SF and CDW QLRO, as a function of $U/t$ and $\Omega_0/U$ for $\delta=0$. (c,d) correspond to $U/t_1=30$, with the lines corresponding to $\delta=0$ and the triangles and diamonds corresponding to $\delta=U/2$. (c) shows $n_{\text{pair}}$ (blue line and triangles) and $m=M/N$ (red dashed line and diamonds) while the inset corresponds to a zoom-in on the polarized regime. (d) shows the ground-state energy (black line and triangles) and the energy of the fully paired (blue dotted line) and fully polarized (red dashed line) state in the absence of the hopping, where $|E_{\rm pgs}|=NU/2$ is the unit of energy. All results are for $N=L=100$.}
\label{fig:Halffillingresults} 
\end{figure*}

In this subsection, we show results obtained by DMRG simulations of Eq.~\ref{eq:BasicHubbardHamiltonian} for $t_2=0$ and half filling. All results are calculated for $N=100$ and $L=100$. In Fig.~\ref{fig:Halffillingresults}(a,b) we plot $\lambda_{P,0}$ and $\lambda_{C,0}$ as functions of the interaction strength $U/t$ and the Rabi coupling $\Omega_0/U$ for $\delta=0$. While the CDW QLRO is always large for $\Omega_0<U$, a clear maximum in $\lambda_{P,0}$ emerges when $\Omega_0$ is slightly smaller than $U$ (although the CDW QLRO is still dominant), particularly for strong interactions. The area in which this enhancement is observed is very small, however, as a transition to an almost polarized band insulator (in which both the SF and CDW off-diagonal correlations decay exponentially) takes place at $\Omega_0 \approx U$. The transition to the band insulator takes place at smaller values of $\Omega_0/U$ for smaller interactions $U$. 

\begin{figure}
\centering
\includegraphics[width=0.9\linewidth]{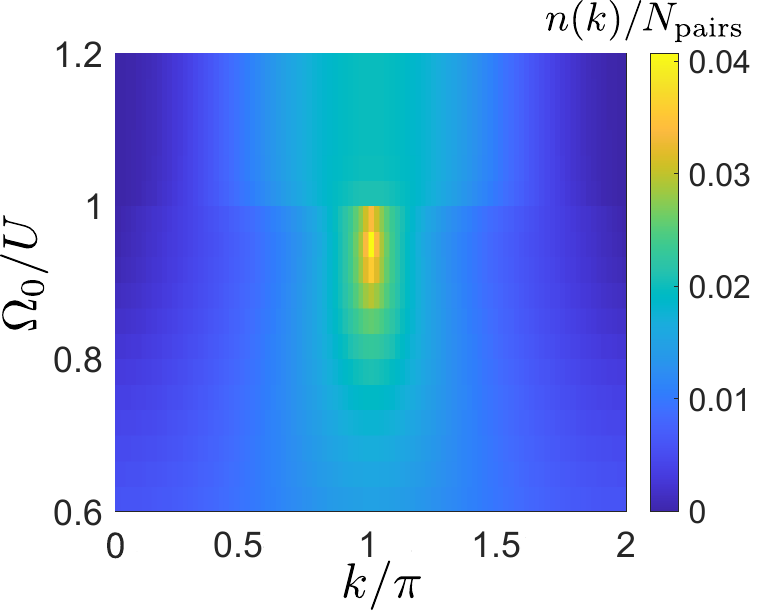}
\caption{The pair momentum distribution at half filling ($N=L=100$) is plotted as a function of $\Omega_0/U$ ($\delta=0$) for $U=30 t_1,t_2=0$.} 
\label{fig:momentumdistribution} 
\end{figure}

$n_{\text{pair}}$ and $m=M/N$ (Fig.~\ref{fig:Halffillingresults}(c)) as well as the energy (Fig.~\ref{fig:Halffillingresults}(d)) as functions of $\sqrt{\delta^2+\Omega_0^2}$ for $\delta=0$ and $\delta=\frac{1}{2}U$ are also shown. Unlike the symmetric case ($t_1=t_2$) where the Rabi coupling is equivalent to a Zeeman field and $[\hat{H}_\theta,\hat{n}_{\pm,j}]=0$ as $\gamma_\theta=0$, the magnetization is not a good quantum number for $t_1 \neq t_2$ and $\Omega_0 \neq 0$, i.e., $[\hat{H}_\theta,\hat{n}_{\pm,j}] \neq 0$. In the former case, the system becomes fully polarized, while it only asymptotically approaches 1 after the transition in the latter case (see the inset of Fig.~\ref{fig:Halffillingresults}(c)) due to the presence of spin fluctuations. The behavior as a function of  $\sqrt{\delta^2+\Omega_0^2}$ is very similar for any combination of $\Omega_0,\delta$ (as long as $\Omega_0\neq0$). Finally we plot the pair momentum distribution in Fig.~\ref{fig:momentumdistribution}. Before the transition a clear peak due to the presence of SF QLRO is visible. See the next two sections for a more detailed explanation of both the momentum distribution (e.g. why the peak is shifted from $k=0$ to $k=\pi$) and the physics in the paired regime in general. 

\subsection{Effective Hamiltonian for large attractive interactions}
\label{sec:EffectiveHamiltonian}

In the limit of strong attractive interactions, an effective Hamiltonian in terms of the fermion-fermion composite bosons can be derived using second-order degenerate perturbation theory. The perturbation starts from the ground state of the Hamiltonian without hopping terms, given by $N/2$ pairs with an energy $E_{\rm pgs}= - NU/2 $ considering the hopping terms as a perturbation. Note that, for $U<\sqrt{\delta^2+\Omega_0^2}$, the state consisting of $N$ spin-polarized fermions becomes the ground state, and the perturbation definitively breaks down. This approximately corresponds to the transition point identified in section III.B and the energies plotted in Fig.~\ref{fig:Halffillingresults}(d) correspond to $-NU/2 $ (blue dotted line) before the transition, while the energy after the transition corresponds to $-\frac{N}{2}\sqrt{\delta^2+\Omega_0^2}$ (red dashed line). The alignment of the energies shown in Fig.~\ref{fig:Halffillingresults}(d) and the high degree of pairing displayed in Fig.~\ref{fig:Halffillingresults}(c) for $\sqrt{\delta^2+\Omega_0^2}<U$ indicates that the perturbation theory should describe the system well for large values of $U$.

Carrying out the perturbation (see the appendix for details), an effective Hamiltonian given in terms of the hard-core bosons $\hat{a}_j= \hat{c}_{j,+}\hat{c}_{j,-}$ can be found as
\begin{align}
\hat{H}_{\text{eff}} &= -t_{\text{eff}} \sum_{j} (\hat{a}_j^\dagger \hat{a}_{j+1} +H.C.) \nonumber \\
&+V_{\text{eff}} \sum_{j}  \hat{n}_{j} \hat{n}_{j+1}-\mu_{\text{eff}} \sum_{j}  \hat{n}_{j},
\end{align}
where
\begin{align}
t_{\text{eff}}=\frac{1}{U} \left[2t_1 t_2+\frac{\Omega_0^2}{2(\Omega_0^2+\delta^2)}(t_1-t_2)^2 \left(1-\frac{1}{1-\tilde{\Omega}^2}\right)\right]
\end{align}
and
\begin{align}
V_{\text{eff}}&=\mu_{\text{eff}}= \nonumber \\
&\!\!\! \frac{1}{U} \left[2\left(t_1^2+ t_2^2\right)+\frac{\Omega_0^2}{\Omega_0^2+\delta^2}(t_1-t_2)^2 \left(-1+\frac{1}{1-\tilde{\Omega}^2}\right)\right]
\end{align}
with $\tilde{\Omega}=\sqrt{\Omega_0^2+\delta^2}/U$.

The hardcore bosonic model can be equivalently described by an XXZ model, utilizing the mapping $\hat{a}^\dagger \rightarrow \hat{S}^{+}$, $\hat{a} \rightarrow \hat{S}^{-}$ and $\hat{n} \rightarrow \hat{S}^z+\frac{1}{2}$, which results in
\begin{align}
\hat{H}_{\mathrm{eff}}
\!=\! -J \! \left[\sum_{j} (\hat{S}^{x}_j\hat{S}^{x}_{j+1}+\hat{S}^{y}_j\hat{S}^{y}_{j+1})
\!-\!\Delta\! \sum_{j}\left(\hat{S}^z_{j}\hat{S}^z_{j+1} \!-\!\frac{1}{4}\right)\right]
,
\end{align}
where $J=2 t_\text{eff}$ and $\Delta =\frac{V_\text{eff}}{2t_\text{eff}}$.
Note that the model can be equivalently described by the antiferromagnetic model by applying the unitary transformation $-\hat{H}(J,-\Delta) = \hat{U}^\dagger \hat{H}(J,\Delta) \hat{U}$ and $(\Delta,J) \rightarrow (-\Delta,-J)$, where $\hat{U} = e^{-i \sum_{j=1}^M \hat{S}^z_j}$  \cite{Langari1998}. For $t_1=t_2$ the model corresponds to the SU(2)-symmetric Heisenberg model for any values of $\delta$ and $\Omega_0$. 
The transverse and longitudinal spin correlations, respectively, correspond to the SF and CDW correlations in the full model. 

For $\Omega_0=0$ the above Hamiltonian corresponds to earlier derivations~\cite{Fath1995,Cazalilla2005}. In this case $t_2=0$ (or equivalently $t_1=0$) maps to the pure Ising model. Finite asymmetry interpolates between the two limits, with the SU(2) symmetry being slowly restored as $t_2 \rightarrow t_1$. A Kosterlitz-Thouless (KT) type transition takes place at $\Delta=1$ at half filling, with long-range CDW (antiferromagnetic order in the spin chain) present for $\Delta >1$ in the thermodynamic limit \cite{Cloizeaux1966,Langari1998,Fath1995,Cazalilla2005}. Due to the exponentially slow opening of the gap, however, QLRO is still observed for relatively large system sizes even at half filling.

\begin{figure}
\centering
\includegraphics[width=0.9\linewidth]{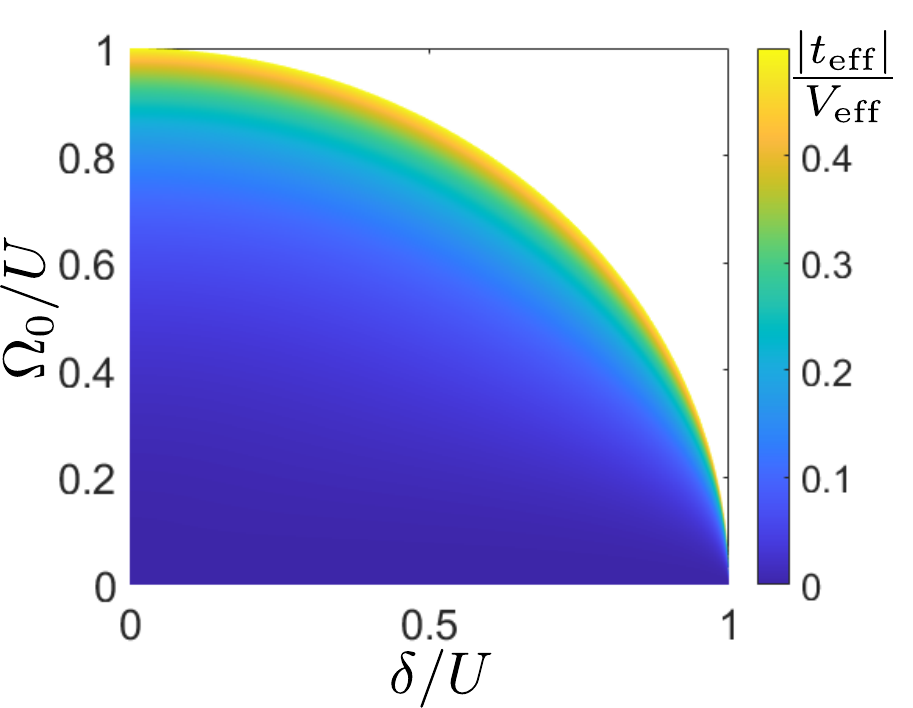}
\caption{The ratio $\frac{|t_{\text{eff}}|}{V_{\text{eff}}}$ as a function of $\delta$ and $\Omega_0$ for $t_2=0$. The white space corresponds to $\sqrt{\Omega_0^2+\delta^2}>U$ where the perturbation is no longer applicable.} 
\label{fig:ratiodeltaomega} 
\end{figure}

Similar observations can be made for the finite $\Omega_0$ case. The sign of $t_{\text{eff}}$ changes when $\sqrt{\Omega_0^2+\delta^2}$ is larger than some value, resulting in a change of sign for both $J$ and $\Delta$ which means that the model corresponds directly to the antiferromagnetic XXZ model. In the limit of $\sqrt{\Omega_0^2+\delta^2}=U$ ($\tilde{\Omega}=1$), $\Delta=-1$ so that the SU(2) symmetry is restored for any combination of $t_1$ and $t_2$. This corresponds to the maximum possible value of the ratio between the effective hopping $t_{\rm eff}$ and repulsive interaction $V_{\rm eff}$ and leads to the same degree of superfluidity as the $t_1=t_2$ case. The effective model breaks down in this limit, however, as this is the exact point at which the energy of the polarized state becomes smaller than that of the paired state. The exact point at which SU(2) symmetry is restored can therefore not be observed in the original FHM. The ratio $\frac{|t_{\text{eff}}|}{V_{\text{eff}}}$ for $t_2=0$ is plotted as a function of $\delta$ and $\Omega_0$ in Fig.~\ref{fig:ratiodeltaomega}. This explains why the results for finite $\delta$ and $\delta=0$ in section \ref{sec:DMRGresulthalffilling} are so similar, in both cases the resonant enhancement is observed as long as $\Omega_0 \neq 0$. From this, we conclude that there is a resonant enhancement of the superfluidity for $\sqrt{\Omega_0^2+\delta^2} \rightarrow U$ as observed in Fig.~\ref{fig:Halffillingresults}(a). $\tilde{\Omega}=\sqrt{\Omega_0^2+\delta^2}/U$ determines the importance of the Rabi contribution to the effective hopping with $\tilde{\Omega}=1$ corresponding to the resonant limit.

The change of sign for $t_{\rm eff}$ in the effective model shifts the energy bands and the lowest energy is therefore expected at $k=\pi$ rather than at $k=0$. From the pair momentum distribution plotted in Fig.~\ref{fig:momentumdistribution} (based on the DMRG calculations of the full FHM), we see that the peak is indeed observed at $k=\pi$.

\subsection{Schematic phase diagram and finite-size scaling}
\label{sec:SchematicPhaseDiagram}

Owing to the presence of the U(1) symmetry breaking for the individual components $\zeta=+$ and $-$ in the Hamiltonian, the system will always have pair-breaking fluctuations. This cannot be captured by the effective model, which describes a definite, conserved number of fermion pairs which leads to an XXZ model with a well-defined and finite spin gap. However, the contribution beyond the effective theory is relatively less important in the regime of applicability of the perturbation theory, which explained the numerical results at half filling fairly well as discussed in the previous section. We therefore expect to be able to understand the important physical properties in the paired regime from the effective model, but will also confirm by numerical calculations of the full model.

\begin{figure}
\centering
\includegraphics[width=0.9\linewidth]{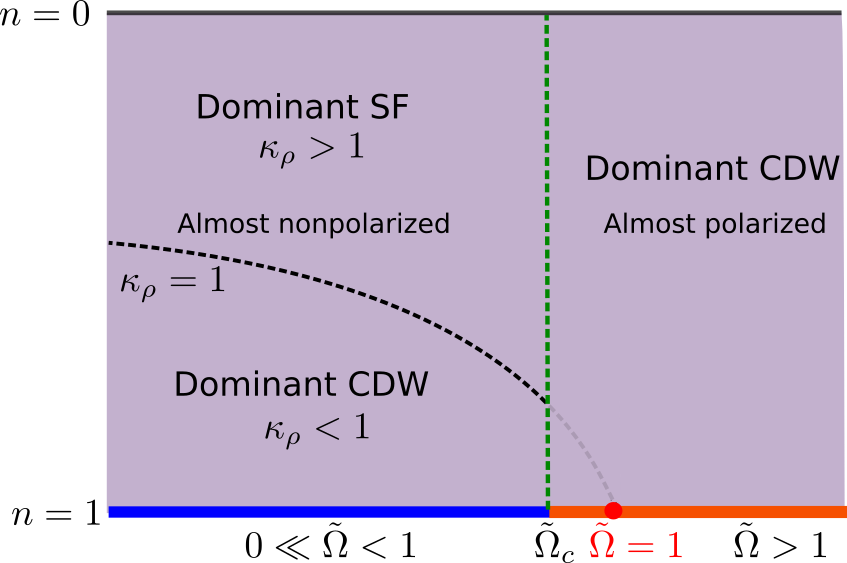}
\caption{Schematic phase diagram as a function of the number of particles per site $n$ and $\tilde{\Omega}$ for $t_2=0$ and fixed large attractive interaction $U$, when $\tilde{\Omega}$ is not too close to zero. For $\tilde{\Omega}<\tilde{\Omega}_c$ at half filling the system displays CDW long-range order (thick blue line), with a transition to a polarized band insulator (thick orange line) taking place at a critical value. For $n<1$ and $\tilde{\Omega}<\tilde{\Omega}_c$ the system corresponds to a Luttinger liquid (LL). The dominant QLRO in the LL depends on the filling $n$ and $\tilde{\Omega}$ and is separated by the line $\kappa_\rho=1$. For $n<1$ and $\tilde{\Omega}>1$ the system becomes mostly polarized and the CDW QLRO is dominant.}
\label{fig:SchematicPhaseDiagram} 
\end{figure}

The one-dimensional XXZ model can be solved exactly utilizing the Bethe ansatz approach~\cite{Yang1966,Giamarchi2003}. As the magnetization in the XXZ model is connected to the number of particles per site in the Hubbard model, we can utilize the results and formulas presented in \cite{Giamarchi2003} to understand the phase diagram of the effective model.  Based on these prior results and numerical evaluations of the full model, the $n$ vs $\tilde{\Omega}=\sqrt{\Omega_0^2+\delta^2}/U$ schematic phase diagram of the RSFHM with $t_2=0$ at constant (relatively large) $U$ is given in Fig.~\ref{fig:SchematicPhaseDiagram}, where $n=N/L$ is the number of particles per site.

For $n=1$ (half filling), the system displays long-range CDW order for any $\tilde{\Omega}<\tilde{\Omega}_c$ according to the effective model. Away from half filling, however, the effective model corresponds to a Luttinger liquid. In the Luttinger liquid regime, the decay of the two correlation functions can be expressed as functions of $r=|i-j|$ \cite{Giamarchi2003,Barbiero2010}
\begin{align}
C(r) & \approx -\frac{\kappa_\rho}{\pi^2 r^2}+A \frac{\cos(2 k_F r)}{r^{\kappa_\rho+\kappa_\sigma}}
\label{eq:LLC} \\
P(r) & \approx \frac{B}{r^{1/ \kappa_\rho+\kappa_\sigma}},
\label{eq:LLP}
\end{align}
where $\kappa_\rho$ and $\kappa_\sigma$ are, respectively, the Luttinger parameters for the charge and spin degrees of freedom and $k_F$ corresponds to the Fermi momentum. Within the effective theory corresponding to the paired regime, the XXZ model is spin-gapped, and we can consider $\kappa_\sigma=0$. The dominant QLRO can then be determined by calculating the Luttinger charge parameter $\kappa_\rho$. The shape of the $\kappa_\rho=1$ line in Fig.~\ref{fig:SchematicPhaseDiagram}, which separates the two regimes of dominant SF or CDW QLRO in the system, is based on numerically solving the Bethe ansatz equations. The Luttinger parameter can also be approximately obtained from the numerical DMRG results in the full model as \cite{Sandvik2004} $\kappa_{\rho} = C(k=2 \pi/L)L/2$ and is expected to converge to a constant with increasing system size in the regime of applicability of the Luttinger theory. The SF QLRO for the Luttinger liquid corresponds to a standard SF (i.e. a single superfluid peak in the pair momentum distribution), and the resonantly enhanced superfluid region described by the effective model is therefore not of the FFLO-type.

\begin{figure*}
\centering
\includegraphics[width=0.7\linewidth]{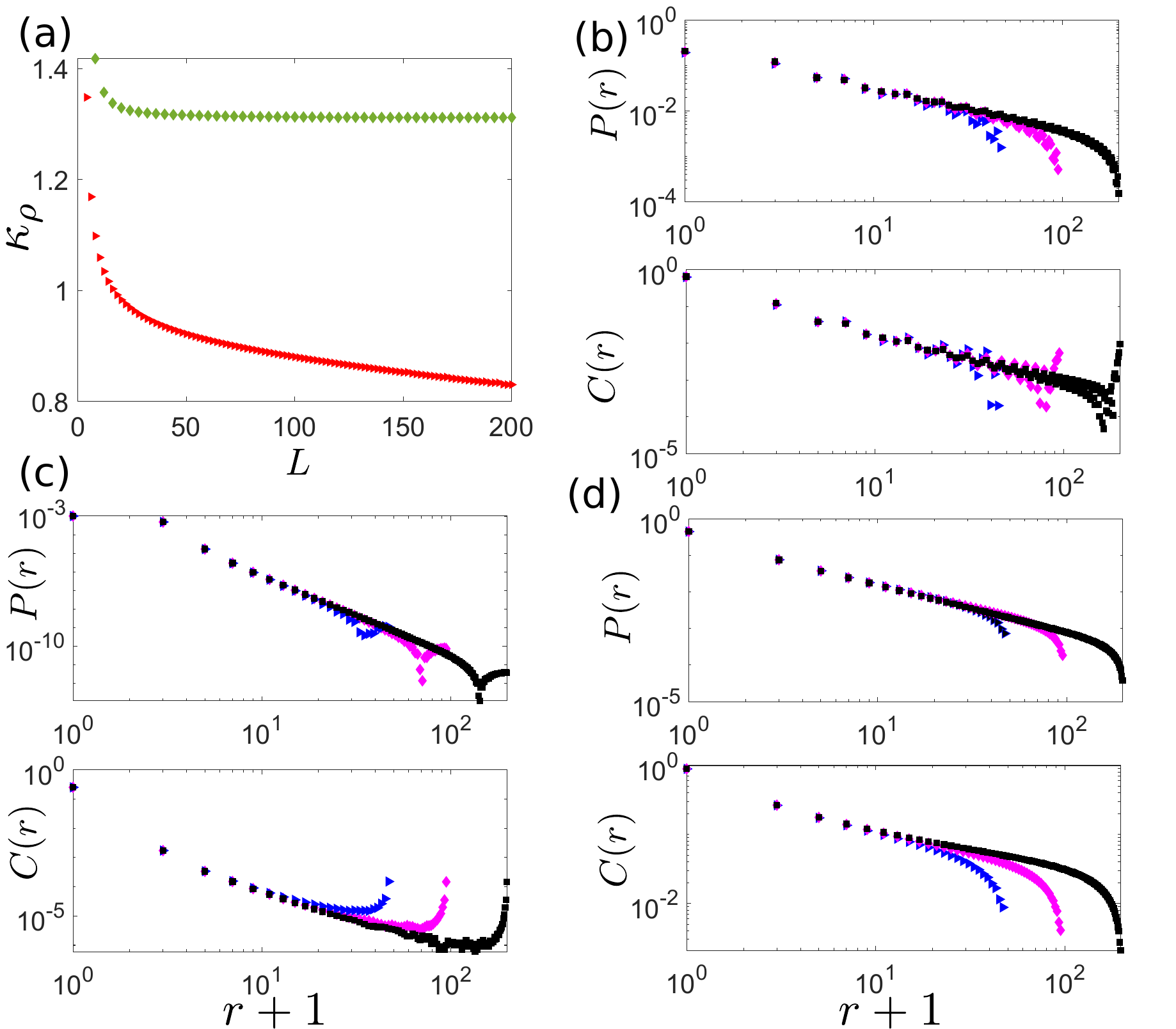}
\caption{ All plots correspond to $t_2=0$, $U=30 t_1$. (a) shows the Luttinger parameter as a function of system size at $\tilde{\Omega}=0.93$ for $n=1/2$ (green diamonds) and $n=1$ (red triangles). (b), (c) and (d) show $P(r)$ and $C(r)$ for  $\tilde{\Omega}=0.93,n=1/2$, $\tilde{\Omega}=1.13,n=1/2$ and $\tilde{\Omega}=0.93,n=1$, respectively. The system sizes are given by $L=48$ (blue triangles), $L=96$ (magenta diamonds) and $L=200$ (black squares).}
\label{fig:finitesizescaling} 
\end{figure*}

At $\tilde{\Omega} \gtrsim 1$, the effective model breaks down as the system becomes mostly polarized, requiring analyses of the full model for understanding the physics. In order to understand this regime and confirm the predictions of the effective model in the regime of its applicability, we consider the decay of the correlation functions with increasing $r=|i-j|$ in the full system. In Fig.~\ref{fig:finitesizescaling}(a), we show the Luttinger parameter $\kappa_\rho$, which is closely related to the algebraic decay of the correlation functions through Eqs.~(\ref{eq:LLC}) and (\ref{eq:LLP}), as a function of the system size (with a maximum of $L=200$) for the half filling ($n=1$) and quarter filling ($n=1/2$) cases at $U=30 t_1$ and $\Omega_0=28 t_1$ ($\tilde{\Omega}=0.93$), which is in the resonant region. Figure~\ref{fig:finitesizescaling}(b-d) show three representative parameter sets of the correlation functions as functions of $r=|i-j|$, which respectively correspond to the cases at half filling ($n=1$) and at quarter filling ($n=1/2$) in the resonant regime $\Omega_0=28 t_1$ ($\tilde{\Omega}=0.93$), and the one at quarter filling ($n=1/2$) in the polarized regime  $\Omega_0=34 t_1$ ($\tilde{\Omega}=1.13$). In order to minimize the boundary effects, we consider $P(r)=P(\frac{L+r}{2},\frac{L-r}{2})$ and $C(r)=C(\frac{L+r}{2},\frac{L-r}{2})$. In general the boundary effects start to come into play as $r>L/2$ (see Fig.~\ref{fig:finitesizescaling}(b-d)) after which the behavior changes drastically. Indeed the calculations at larger system size clearly shows that this change is a boundary effect.

For quarter filling in the resonant regime, we clearly see in Fig.~\ref{fig:finitesizescaling}(a) that the Luttinger parameter is well converged to $\kappa_\rho \approx 1.3$ at $L>50$. This is consistent with the fact that  both the SF and CDW correlations display a power-law decay with $P(r) \propto r^{-0.84}$ and $C(r) \propto r^{-1.2}$ as shown in Fig.~\ref{fig:finitesizescaling}(b). Indeed, these exponents are close to those predicted by the Luttinger theory, which is respectively $-0.76$ and $-1.3$. The actual decay exponents are slightly biased towards CDW QLRO compared to those expected from the Luttinger liquid, which is likely due to the presence of finite polarization in the full model. In the polarized regime, both $P(r)$ and $C(r)$ also display a power-law decay (Fig.~\ref{fig:finitesizescaling}(c)), implying that there is no phase transition between the two regimes. While the pairing fraction in the polarized regime never becomes zero, it asymptotically approaches zero as $\tilde{\Omega}$ increases and the power law decay for the correlations is very fast $P(r) \propto r^{-\eta}~(\eta \approx 4)$. This is in stark contrast to the resonant regime where the SF correlations are dominant at quarter filling and we can therefore clearly distinguish the SF properties of these two regimes. For half filling, both correlation functions display a power-law decay (Fig.~\ref{fig:finitesizescaling}(d)), although a slight flattening of the $C(r)$ curve is visible for large $r$ at $L=200$. The Luttinger parameter decreases with the system size (Fig.~\ref{fig:finitesizescaling}(a)), which suggests that the system is in fact not a Luttinger liquid. This is consistent with the analysis of the effective theory where a KT transition to a long-range ordered CDW phase takes place at the resonant point and long-range CDW is expected for $\tilde{\Omega}<1$ at half filling. We therefore expect that $P(r)$ would start to show exponential decay for larger system sizes and that $C(r)$ would become constant. Note that this means the resonant regime will display SF-like behavior even at half filling for system sizes relevant in most cold atom lattice experiments. At half filling, there is a phase transition to a band insulator as mentioned in section \ref{sec:DMRGresulthalffilling}.

\subsection{SF and CDW order as a function of the particle number per site}
\label{sec:SuperfluidCDWfilling}

From the effective model, we expect the values of  $\tilde{\Omega}$ where superfluidity is dominant to be broadened as $n$ is lowered. However, observing this in the DMRG calculations is difficult because the low-energy states are energetically closer the further one goes from the resonant limit and half filling. Instead, we focus on the resonantly enhanced region, as the DMRG method gives well-converged results in this regime, and investigate the CDW and SF QLRO as a function of the particle number per site $n$, corresponding to vertical lines within the non-polarized regime of the diagram given by Fig.~\ref{fig:SchematicPhaseDiagram}.

\begin{figure*}
\centering
\includegraphics[width=1\linewidth]{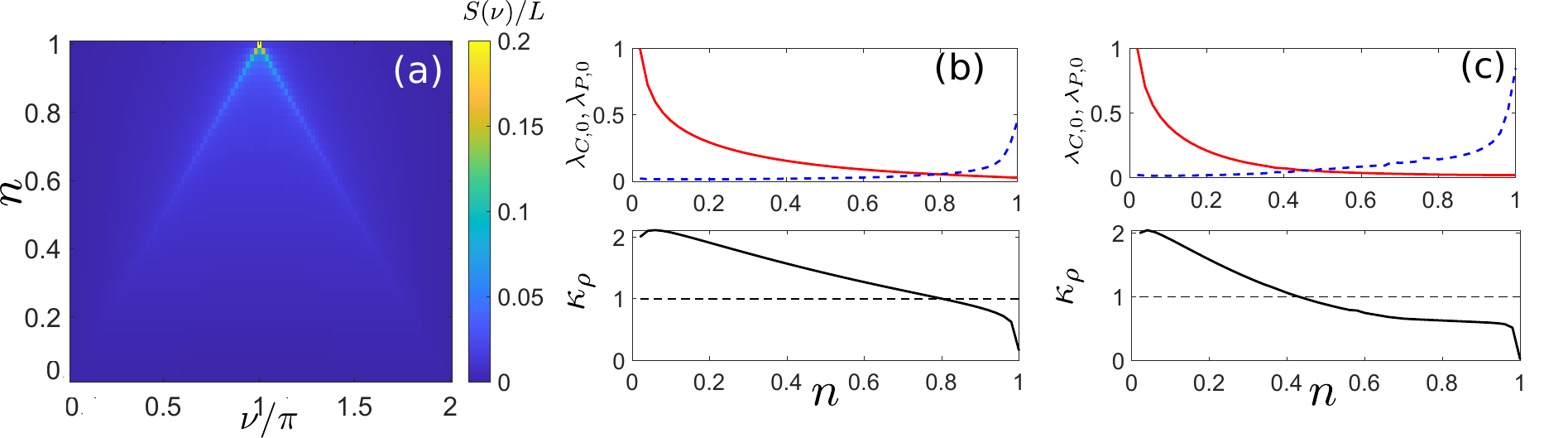}
\caption{The ground state properties are plotted by varying the particle number per site $n=N/L$, where $\delta =0$. (a) shows $S(\nu)/L$ as a function of $n$, where  $U=30t_1$ and $\Omega_0=25t_1$.  (b) and (c) show $\lambda_{P,0}$ (red full line), $\lambda_{C,0}$ (blue dashed line) and $\kappa_\rho$ (black full line) as a function of the filling ratio $n$ for  $U=30t_1$, $\Omega_0=25t_1$, and $U=5t_1,\Omega_0=2.8t_1$ respectively.} 
\label{fig:functionoffillingratio} 
\end{figure*}

To investigate the dominant QLRO as a function of $n$, we compute the Luttinger parameter $\kappa_\rho$ as a function of $n$ for $t_2=0$ in the strongly-interacting region $U=30t_1$, $\Omega_0=25t_1$, $\delta=0$ ($\tilde{\Omega}=0.83$) where the effective model is expected to work well and for more moderate interactions $U=5t_1$, $\Omega_0=2.8t_1$, $\delta=0$ ($\tilde{\Omega}=0.56$). A smaller value of $\tilde{\Omega}$ is chosen in the latter case as the polarization becomes dominant at smaller $\tilde{\Omega}$ for smaller values of $U$. From Fig.~\ref{fig:functionoffillingratio}(a), in which $S(\nu)/L$ is shown, it is clear that the single $\pi$-peak at half filling for the structure factor splits into two symmetric peaks around $\nu=\pi$ and that their absolute values decrease as $n$ is lowered. In the upper panels of Fig.~\ref{fig:functionoffillingratio}(b,c), $\lambda_{P,0}$ and $\lambda_{C,0}$ are plotted. $\lambda_{P,0}$ remains bigger than $\lambda_{C,0}$ for a relatively large $n$ region; $0 < n \lesssim 0.8$ in the case of $U=30t_1$ and $\Omega_0=25t_1$ ($0 < n \lesssim 0.5$ in the case of $U=5,\Omega_0=2.8$).
In Fig.~\ref{fig:functionoffillingratio}(b,c), we also plot the Luttinger parameter $\kappa_\rho$ in the lower panels and show that these crossing points of $\lambda_{P,0}$ and $\lambda_{C,0}$ correspond to the point where $\kappa_\rho$ crosses 1 signifying the change in dominant QLRO. Overall this implies that for lower $n$ the SF QLRO is dominant. These results correspond to a vertical slice of the schematic phase diagram in Fig.~\ref{fig:SchematicPhaseDiagram} with $\tilde{\Omega}=0.83$ and $0.56$, respectively, where both are below the value at which polarization becomes dominant, which is smaller for the smaller value of $U$.

\section{Conclusion}                                                         
\label{sec:summary}
 
We have shown that superfluidity can appear by tuning the Rabi coupling in a spin-dependent Fermi-Hubbard model for any combination of hoppings. In particular, this still holds when one component is immobile and the superfluidity is entirely absent without the hybridization introduced by the Rabi coupling. The maximum enhancement of superfluidity happens when the Rabi coupling is slightly smaller than the interaction strength, just before the system starts to strongly polarize.  A slightly doped system (compared to the commensurate half-filling case) displays a sharply resonant enhancement, while larger doping allows enhancement to be observed at somewhat smaller Rabi couplings.  

While our results focus on the one-dimensional case where exact numerical results can be obtained by use of the DMRG method, the rotated Hamiltonian introduced in Eq.~\eqref{eq:RotatedHubbardHamiltonian} and the effective model we derive at large attractive interactions are easily generalized to higher dimensions. An investigation into the higher-dimensional case would therefore be an interesting future direction. Since the effect of finite temperature should be taken into account for a quantitative comparison with experiments, an extension to finite-temperature systems is a natural continuation of our research as well.

\begin{acknowledgments}
We would like to thank S. Goto for helpful discussions with respect to the numerical DMRG calculations done in this paper. The calculations were performed utilizing the ITensor C++ library \cite{Fishman2020}. This work was financially supported by JST CREST (Grant No.\ JPMJCR1673), JST FOREST (Grant No.\ JPMJFR202T), MEXT Q-LEAP (Grant No.\ JPMXS0118069021), and KAKENHI from Japan Society for Promotion of Science (Grant Nos.\ JP18H05228, JP21H01014, and JP21K13855).
\end{acknowledgments}

\appendix
\section{Derivation of the effective model at large attractive interactions}
\label{sec:appendix}

In order to carry out the derivation of the effective Hamiltonian, let us write the rotated Hamiltonian given by Eq. (\ref{eq:RotatedHubbardHamiltonian}) as 
\begin{align}
\hat{H}_{\theta} &= \hat{H}_0+\hat{H}_{1} \\
\hat{H}_1&=\hat{T}+\hat{\Gamma}_{+}+\hat{\Gamma}_{-}\\
\hat{H}_0& = \sum_{j} \Big[ (-\mu+\delta_{\theta})\hat{n}_{+,j}+ (-\mu-\delta_{\theta})\hat{n}_{-,j} \Big]
\nonumber\\
&\,\,\,\, -U\sum_{j}\hat{n}_{+,j}\hat{n}_{-,j} ,
\end{align}
where 
\begin{align}
\hat{T} &= \sum_{\zeta=+,-}\sum_{j} \Big( -t_{\theta,\zeta}  \hat{c}_{\zeta,j}^\dagger \hat{c}_{\zeta,j+1}+H.C. \Big)\\
\hat{\Gamma}_{+}&=\gamma_\theta \sum_{j}\left(\hat{c}_{+,j}^\dagger \hat{c}_{-,j+1}+\hat{c}_{+,j+1}^\dagger \hat{c}_{-,j}\right) \\
\hat{\Gamma}_{-}&=\gamma_\theta\sum_{j}\left( \hat{c}_{-,j}^\dagger \hat{c}_{+,j+1}+\hat{c}_{-,j+1}^\dagger \hat{c}_{+,j}\right).
\end{align}

As a starting point for deriving the effective Hamiltonian, we consider the Hamiltonian $\hat{H}_0$. For $U>\sqrt{\delta^2+\Omega_0^2}$ the manifold of degenerate ground states $|\psi_{P,j}\rangle \in \mathcal{P}$ is given by all combinations of $N/2$ pairs (for example $\mathcal{P}=\{|220\rangle,|202\rangle,|022\rangle\}$  for $N=4,M=3$). The energy of these states is $-N/2 U-N\mu$. The effective Hamiltonian is derived by considering the hopping terms as a perturbation. The excited eigenstates correspond to the complement of the ground-state manifold and are denoted by $|\psi_{Q,j}\rangle \in \mathcal{Q}$. Applying the hopping terms to the ground states always creates states in the complement and the first order contributions of the perturbation is therefore zero, $\sum_{j,k}\langle \psi_{P,j}|\hat{H}_1|\psi_{P,k}\rangle | \psi_{P,j}   \rangle \langle \psi_{P,k} |=0$. The effective Hamiltonian is then given by the second-order term
\begin{align}
&\hat{H}_{\text{eff}}=\sum_{i,j,k} \frac{1}{E_0-E_j} | \psi_{P,i} \rangle \langle \psi_{P,i} | \hat{T}+\hat{\Gamma}_{+}+\hat{\Gamma}_{-} | \psi_{Q,j} \rangle \times \nonumber \\
&\langle \psi_{Q,j} |  \hat{T}+\hat{\Gamma}_{+}+\hat{\Gamma}_{-} | \psi_{P,k} \rangle \langle \psi_{P,k} | .
\end{align}
Note that only terms which can connect two states from the ground-state manifold to the same state in the complement can have any contribution. $\hat{\Gamma}_{+}$ connects to a state with one broken pair and $N_{+}=N/2+2$ with a corresponding energy $E_j=\frac{-N}{2}U+U-N \mu+\sqrt{\Omega_0^2+\delta^2}$, whereas $\hat{\Gamma}_{-}$ connects to a state with a broken pair and $N_{-}=N/2+2$ with a corresponding energy  $E_j=\frac{-N}{2}U+U-N \mu-\sqrt{\Omega_0^2+\delta^2}$. $\hat{T}$ connects to a state with a broken pair and $N_{+}=N_{-}=N/2$ with a corresponding energy $E_j=\frac{-N}{2}U+U-N \mu$. This means that  $\langle \psi_{P,i}| \hat{T} | \psi_{Q,j} \rangle \langle \psi_{Q,j} | \hat{\Gamma}_{\pm} | \psi_{P,k} \rangle$ will have no contribution. Similarly 
$\langle \psi_{P,i}| \hat{\Gamma}_{\pm} | \psi_{Q,j} \rangle \langle \psi_{Q,j} | \hat{\Gamma}_{\pm} | \psi_{P,k} \rangle$ has no contribution and we are left with three terms which can be written like
\begin{align}
 \hat{H}_{\text{eff},T}=\frac{-1}{U} \sum_{i,k}\langle \psi_{P,i} | \hat{T}^2 | \psi_{P,k} \rangle | \psi_{P,i} \rangle  \langle \psi_{P,k} |,
 \label{eq:case1}
\end{align} 
\begin{align}
 \hat{H}_{\text{eff},+-}= \frac{-1}{U-\sqrt{\Omega_0^2+\delta^2}} \sum_{i,k} \langle \psi_{P,i} |\hat{\Gamma}_{+}\hat{\Gamma}_{-} | \psi_{P,k} \rangle | \psi_{P,i} \rangle \langle \psi_{P,k} |
,
  \label{eq:case2}
\end{align} 
and
\begin{align}
 \hat{H}_{\text{eff},-+}= \frac{-1}{U+\sqrt{\Omega_0^2+\delta^2}}  \sum_{i,k} \langle \psi_{P,i} |\hat{\Gamma}_{-}\hat{\Gamma}_{+} | \psi_{P,k} \rangle | \psi_{P,i} \rangle \langle \psi_{P,k} |.
\label{eq:case3}
\end{align} 
Here we have utilized that $E_j$ is constant for any value of the states in the complement that can be connected to by these terms. This allows us to pull it out of the sum after which we see that the remaining sum corresponds to a matrix product leading to the above equations. Note that the first term corresponds to a perturbation in the absence of the Rabi-coupling term which has already been carried out in, e.g.,~\cite{Fath1995,Cazalilla2005}.

In order to derive the effective Hamiltonian, we need to evaluate the matrix elements. As these only connect to neighboring pairs we can do this by considering the effect on the two-site states $|2 0 \rangle$ and $|0 2 \rangle$ (corresponding to the $j$ and $j+1$ sites). Applying the local hopping elements twice to the states corresponding to Eq. (\ref{eq:case1}), the non-zero terms can be arranged as 
\begin{align*}
t_{\theta,+} t_{\theta,-} \hat{c}_{+,j}^\dagger \hat{c}_{+,j+1} \hat{c}_{-,j}^\dagger \hat{c}_{-,j+1} |0 2\rangle =t_{\theta,+} t_{\theta,-} |2 0\rangle\\
t_{\theta,-} t_{\theta,+} \hat{c}_{-,j}^\dagger \hat{c}_{-,j+1} \hat{c}_{+,j}^\dagger \hat{c}_{+,j+1} |0 2\rangle = t_{\theta,-} t_{\theta,+} |2 0\rangle\\
t_{\theta,+} t_{\theta,+} \hat{c}_{+,j+1}^\dagger \hat{c}_{+,j} \hat{c}_{+,j}^\dagger \hat{c}_{+,j+1} |0 2\rangle =t_{\theta,+} t_{\theta,+} |0 2 \rangle \\
t_{\theta,-} t_{\theta,-} \hat{c}_{-,j+1}^\dagger \hat{c}_{-,j} \hat{c}_{-,j}^\dagger \hat{c}_{-,j+1} |0 2\rangle =t_{\theta,-} t_{\theta,-} |0 2 \rangle
\end{align*}
and
 \begin{align*}
 t_{\theta,+} t_{\theta,-} \hat{c}_{+,j+1}^\dagger \hat{c}_{+,j} \hat{c}_{-,j+1}^\dagger \hat{c}_{-,j} |2 0\rangle =t_{\theta,+} t_{\theta,-} |0 2\rangle \\
t_{\theta,-} t_{\theta,+} \hat{c}_{-,j+1}^\dagger \hat{c}_{-,j} \hat{c}_{+,j+1}^\dagger \hat{c}_{+,j} |2 0\rangle = t_{\theta,-} t_{\theta,+} |0 2\rangle \\
t_{\theta,+} t_{\theta,+} \hat{c}_{+,j}^\dagger \hat{c}_{+,j+1} \hat{c}_{+,j+1}^\dagger \hat{c}_{+,j} |2 0\rangle =t_{\theta,+} t_{\theta,+} |2 0 \rangle \\
t_{\theta,-} t_{\theta,-} \hat{c}_{-,j}^\dagger \hat{c}_{-,j+1} \hat{c}_{-,j+1}^\dagger \hat{c}_{-,j} |2 0\rangle =t_{\theta,-} t_{\theta,-} |2 0 \rangle .
\end{align*}

These can be expressed in terms of the effective hard-core boson operators $\hat{a}_j=\hat{c}_{+,j} \hat{c}_{-,j}$  and $\hat{n}_j=\hat{a}_j^\dagger \hat{a}_j$. As we only consider states with zero or pair occupation $\hat{n}=\hat{n}_+=\hat{n}_-$ at each site. Utilizing this we can rewrite the sum as that over the effective operators
\begin{eqnarray}
 &&-2\frac{t_{\theta,+}t_{\theta,-}}{U}(\hat{a}_j^\dagger \hat{a}_{j+1} + \hat{a}_{j+1}^\dagger \hat{a}_{j})
 \nonumber\\
 &&+\frac{t_{\theta,+}^2+t_{\theta,-}^2}{U}(2 \hat{n}_j \hat{n}_{j+1}-\hat{n}_j-\hat{n}_{j+1}).
\end{eqnarray}
Here we utilized the fact that $\hat{c}_{\sigma,j} \hat{c}_{\sigma,j}^\dagger =1-\hat{c}_{\sigma,j}^\dagger \hat{c}_{\sigma,j}$ and the fermionic anti-commutation relations for switching operators at different sites. 

The inter-species hopping terms can be arranged as
\begin{align*}
\gamma_{\theta}^2 \hat{c}_{+,j}^\dagger \hat{c}_{-,j+1} \hat{c}_{-,j}^\dagger \hat{c}_{+,j+1} |0 2\rangle =\gamma_{\theta}^2 |2 0\rangle \\
\gamma_{\theta}^2 \hat{c}_{+,j+1}^\dagger \hat{c}_{-,j} \hat{c}_{-,j}^\dagger \hat{c}_{+,j+1} |0 2\rangle = \gamma_{\theta}^2 |0 2\rangle \\
\gamma_{\theta}^2 \hat{c}_{+,j+1}^\dagger \hat{c}_{-,j} \hat{c}_{-,j+1}^\dagger \hat{c}_{+,j} |2 0\rangle = \gamma_{\theta}^2 |0 2\rangle \\
\gamma_{\theta}^2 \hat{c}_{+,j}^\dagger \hat{c}_{-,j+1} \hat{c}_{-,j+1}^\dagger \hat{c}_{+,j} |2 0\rangle = \gamma_{\theta}^2 |2 0\rangle
\end{align*}
for Eq. (\ref{eq:case2}) and
\begin{align*}
\gamma_{\theta}^2 \hat{c}_{-,j}^\dagger \hat{c}_{+,j+1} \hat{c}_{+,j}^\dagger \hat{c}_{-,j+1} |0 2\rangle = \gamma_{\theta}^2 |2 0\rangle \\
\gamma_{\theta}^2 \hat{c}_{-,j+1}^\dagger \hat{c}_{+,j} \hat{c}_{+,j}^\dagger \hat{c}_{-,j+1} |0 2\rangle = \gamma_{\theta}^2 |0 2\rangle \\
\gamma_{\theta}^2 \hat{c}_{-,j+1}^\dagger \hat{c}_{+,j} \hat{c}_{+,j+1}^\dagger \hat{c}_{-,j} |2 0\rangle =  \gamma_{\theta}^2 |0 2\rangle \\
\gamma_{\theta}^2 \hat{c}_{-,j}^\dagger \hat{c}_{+,j+1} \hat{c}_{+,j+1}^\dagger \hat{c}_{-,j} |2 0\rangle =  \gamma_{\theta}^2  |0 2\rangle
\end{align*}
for Eq. (\ref{eq:case3}). The sums can be rewritten in terms of the effective operators, using the same properties as for the intra-species hopping terms
\begin{align}
&\frac{2 U\gamma_{\theta}^2}{U^2-(\Omega_0^2+\delta^2)}\times
\nonumber\\ 
&\left(\hat{a}_j^\dagger \hat{a}_{j+1} +\hat{a}_{j+1}^\dagger \hat{a}_{j}+2\hat{n}_j \hat{n}_{j+1}-\hat{n}_j-\hat{n}_{j+1}\right).
\end{align}
Note that for the kinetic energy terms the fermionic commutation relations resulted in a change of sign compared to the intra-species hopping case. Combining all terms the effective Hamiltonian can be expressed using the hard-core bosonic operator as  
\begin{align}
\hat{H}_{\text{eff}} &= -t_{\text{eff}} \sum_{j} (\hat{a}_j^\dagger \hat{a}_{j+1} +H.C) \nonumber\\
&+V_{\text{eff}} \sum_{j}  \hat{n}_{j} \hat{n}_{j+1}-\mu_{\text{eff}} \sum_{j}  \hat{n}_{j}
\end{align}
with 
\begin{align}
t_{\text{eff}}&=\frac{2}{U}t_{\theta,+} t_{\theta,-} - 2\frac{U\gamma_{\theta}^2}{U^2-(\Omega_0^2+\delta^2)} 
\end{align}
and
\begin{align}
V_{\text{eff}}&=\mu_{\text{eff}}= \frac{2}{U}(t_{\theta,+}^2+t_{\theta,-}^2)+ 4\frac{U\gamma_{\theta}^2}{U^2-(\Omega_0^2+\delta^2)}.
\end{align}
By inserting the values of $t_{\theta,+},t_{\theta,-},\gamma_\theta$ the equations in the main text are obtained. 

\bibliography{library}

\end{document}